# Health Care Expenditures, Financial Stability, and Participation in the Supplemental Nutrition Assistance Program (SNAP)

**Yunhee Chang, Ph.D.**

**University of Mississippi**

**Jinhee Kim, Ph.D.**

**University of Maryland**

**Swarn Chatterjee, Ph.D.**

**University of Georgia**

## Abstract

This paper examines the association between household healthcare expenses and participation in the Supplemental Nutrition Assistance Program (SNAP) when moderated by factors associated with financial stability of households. Using a large longitudinal panel encompassing eight years, this study finds that an inter-temporal increase in out-of-pocket medical expenses increased the likelihood of household SNAP participation in the current period. Financially stable households with precautionary financial assets to cover at least 6 months' worth of household expenses were significantly less likely to participate in SNAP. The low-income households who recently experienced an increase in out-of-pocket medical expenses but had adequate precautionary savings were less likely than similar households who did not have precautionary savings to participate in SNAP. Implications for economists, policy makers, and household finance professionals are discussed.

**Introduction**

The Supplemental Nutrition Assistance Program (SNAP), formerly known as the Food Stamp Program, provides benefit payments to purchase food for households meeting the eligibility criteria. SNAP benefits have been found to help low-income families smooth their consumption (Gundersen and Ziliak 2003) and serve as an economic safety net in the events of negative income shocks. SNAP participation rates have increased in the past decade (Zedlewski and Rader 2005), reaching over 47 million recipients in 2012 (Food and Nutrition Services [FNS], U.S. Department of Agriculture [USDA] 2013). Recent increases in participation have been largely explained by the unemployment rates and the number of people in poverty (Andrews and Smallwood 2012; Klerman and Danielson 2011; Lim 2011). Recent policy modifications at both federal and state levels, such as reductions in certification process and more lenient vehicle exemption, were also found to have led to increases in SNAP participation (Klerman and Danielson 2011).

Financial instability and liquidity constraint of individual households have been associated with SNAP participation (Mabli & Ohls, 2012). Households that experience financial strain were more likely to participate in SNAP (Purtell, Gershoff, and Aber 2012). While the indicators of household income loss such as unemployment, employment changes, and job instability have been associated with SNAP participation (Mabli and Ohls 2012; Yen, Bruce, and Jahns 2012), the impact of unexpected major expenses such as medical bills has rarely been studied in relation to SNAP participation. With increased health care expenditures and out-of-pocket costs, medical expenses have become a major contributor to household financial instability (Collins et al. 2008). On another note, liquidity constraints diminish the financial stability of households (Cox and Jappelli 1993; Grafova 2011; Grenninger et al. 1996). Assets and liquidity may help households in coping with financial shocks without turning to public assistance such as SNAP. However, many households are inadequately prepared to deal with the sudden increases in out-of-pocket medical expenses (Feenberg and Skinner 1994; McIntyre et al. 2006; Nielsen, Garasky, and Chatterjee 2010). Increases in out-of-pocket medical expenses can particularly hurt households that do not have adequate reserves of emergency funds to buffer such financial shocks (Kim and Lyons 2008; Kim, Yoon, and Zurlo 2012).

The purpose of this research is to examine the effect of health care burdens on the SNAP participation of households. This study specifically examines the following three research questions: (1) whether increases in households' out-of-pocket medical expenses are associated with their likelihood of participating in the SNAP, (2) whether households' liquidity constraint is associated with their likelihood of participating in SNAP, and (3) whether the absence of liquidity constraint reduces the association between out-of-pocket medical expenditure and SNAP participation.

**Literature Review**

*Health Care Expenditures and Financial Strain*

Medical expenses have become a major cause of households' financial instability. Many Americans are struggling to pay their medical bills and accumulating large amounts of medical debt. When compared with higher-income households, financial burden of medical expenses was greater for low-income families (Cohen and Kirzinger 2014; Patel, Brown, and Clark 2013) and for the uninsured (Bernard, Johansson, and Fang 2014). Households with special medical needs often experienced high levels of financial strain (Lindley and Mark 2010). About a quarter of those who were uninsured in the previous year were unable to pay their medical bills (Collins et al. 2008). Having private health insurance coverage offered households little protection from financial burden of medical bills due to high premium and out-of-pocket costs (Cohen, Gindi, and Kirzinger 2012).

The highest levels of financial burden of medical cost were found in poor (below the Federal Poverty Line [FPL]) and near-poor families (100-200% FPL) (Cohen et al. 2012). Ketsche, Adams, Wallace, Kannan, and Kannan (2011) examined health care expenditures including health insurance and out-of-pocket health care spending by income group. They found that lower-income families paid a larger share of their incomes on health care than higher-income families did. Out-of-pocket expenditures for low-income families represented a larger proportion of the family income and thus lead to relatively greater financial burden (Witt et al. 2011). Galbraith, Wong, Kim, and Newacheckal (2005) found that lower-income groups reported greater out-of-pocket expenditures per $1,000 income than other income groups. Similarly, Selden (2009) showed that lower-income families were more likely to incur out-of-pocket expenditures exceeding 20% of family income compared to higher income families.

Families with low income, children, and limited or no insurance coverage experienced higher financial burdens of medical care than others (Cohen and Krizinger 2014). With a population of 47 million nonelderly uninsured low-income families bear higher financial risks due to lack of insurance or inadequate health insurance coverage (i.e., underinsurance). Many low-income families have Medicaid and other public health coverage but many of their family members have been uninsured or underinsured in the past. Although health insurance coverage alleviates the burden to some extent, out-of-pocket financial burden for low-income families with children is significantly higher than other income groups (200% FPL or higher) regardless of their health insurance coverage (Cohen and Krizinger 2014). Public insurance programs may have minimal cost sharing but may not cover all of the services that are needed. Although, support for low-income children are available at the state level through the State Children's Health Insurance Programs (SCHIP), evidence suggests that these public insurance programs and the traditional private insurance policies may not eliminate the out of pocket medical costs for low-income families, especially for those with chronic conditions (Lindley and Mark 2010).

Additionally, medical expenses are the leading cause of consumer bankruptcies (Dranove and Millenson 2006) and out-of-pocket medical expenditures play an increasing role in one out of four low-income household bankruptcies (Gross and Notowidigo 2011). Although the Patient Protection and Affordable Care Act of 2010 (ACA) offers opportunities to extend the coverage of many uninsured people, financial burden of health care may continue to affect low-income households for a while. A study that followed the universal medical insurance coverage in Massachusetts found that bankruptcy filings increased in Massachusetts (Badding, Stephenson, and Yeoh 2012) whereas a more recent study showed the broader positive impacts of universal insurance coverage in Massachusetts on credit scores and reduced personal bankruptcies (Mazumder and Miller 2014).

*Health Care Expenditures, Household Financial Instability, and SNAP Participation*

Low-income households have limited monthly budgets and spend a large share of their income on basic needs such as food, housing, and medical expense. The elderly and disabled members from low-income households are at especially higher risk for financial burden due to medical expenses. Additionally, the low-income mothers have a higher likelihood of suffering from chronic diseases and health conditions than other groups (Bombard et al. 2012).

Previous studies have found that health care expenditures are associated with food insecurity because medical expenses can crowd out the households' ability to purchase food (Biros, Hoffman, and Resch 2005; Lee 2013; Patton-Lopez 2012). In many low-income families, poor health conditions forces them to choose between food and medicine, increasing their risk of cutting back on expenses for food, medical expenses, or both (Lee 2013). One study conducted by Nielsen et al. (2010) found that probability of experiencing food insecurity increased as the out-of-pocket medical expenditures increased. These findings indicate that expenditures on medical care may reduce the resources available for food consumption.

While the burden of health care costs can aggravate food insecurity, the reverse may also be true. Negative health outcomes such as chronic and mental health problems and emergency room visits resulting from hunger and food insecurity have well been documented, especially for low-income individuals and families (Biros et al. 2005; Sullivan et al. 2010). Not having enough money for food and health care may deteriorate health and require greater health care costs (Biros et al. 2005; Patel et al. 2012; Sullivan et al. 2010).

Currently, in most states medical care costs are deductible expenses for households with members who are elderly and for households with disabilities in calculating SNAP eligibility and benefits. Eligible households can deduct out-of-pocket medical expenses that are more than $35. The deduction for excessive medical expenses can lead to a substantial increase in SNAP benefits (USDA, 2014). A broad array of medical expenses, including transportation costs to a pharmacy or a doctor's office, over-the-counter drugs, medical supplies, and home renovations to increase accessibility, are eligible for deduction. For eligible seniors or disabled individuals, a claim of $50-$200 in monthly medical expenses can result in a monthly increase of $7-$69 in SNAP benefits. Given the fact that families experiencing food insecurity often enroll in federal food assistance programs such as SNAP (Mammen, Bauer, and Richards 2009; Swanson et al.

2008), it is reasonable to expect that increased financial burden of health care expenses that increase food insecurity could also lead to increased SNAP participation.

*Financial Instability and SNAP Participation*

Households that experienced poverty and financial strains were more likely to participate in SNAP (Purtell et al. 2012). There is a lack of consensus in the literature, however, regarding the issue of measuring household financial stability. Income and employment were often used as measures of financial stability (Mabli and Ohls 2012; Yen et al. 2012) where limited research is available regarding the relationship between assets and SNAP participation. Financial assets can be used to maintain food consumption when households face income volatility. Previous studies have established a number of financial ratio measures such as liquidity constraint, asset inadequacy, and insolvency to assess household financial stability (Bi and Montalto 2004; Choi et al. 2001; DeVaney 2002; Grafova 2011; Harness, Chatterjee, and Finke 2009). Household liquidity constraint was positively associated with financial strain (Grafova 2011; Cox and Jappelli 1993). A recent study found that the effects of household asset holdings and debt burden on food insecurity were separate from the effect of current-period income (Chang, Chatterjee, and Kim 2013). Liquidity constraint might affect the association between medical expense and SNAP participation as any assets and savings can be used to buffer financial stress such as increased medical care expenses.

**Methods**

*Data*

This study used the 2003-2011 Panel Study of Income Dynamics (PSID). The PSID is a longitudinal survey that began in 1968 with a nationally representative sample of over 5,000 households. It currently collects household- and individual-level information from over 9,000 households on various topics on a biennial basis. Our sample was drawn from the 2003, 2005, 2007, 2009, and 2011 data, which cover the period of recent financial crisis and recession.

Previous studies have shown that the health insurance related variables in the PSID compare well with the Medical Expenditure Panel Survey (MEPS) and National Health Interview Survey (Levy, 2007). However, using the PSID provides several advantages for this study that was not available in other datasets such as the MEPS or NHIS. First, the PSID not only identifies SNAP participating households, it is one of a few nationally representative surveys that include detailed information on household assets and liability. Based on the detailed assets and wealth data, we constructed liquidity constraint measures. Second, the PSID's individual questionnaire includes detailed questions on various types of health-related expenditures and health insurance. Third, the longitudinal nature of the data not only allows us to investigate changes over time in household conditions, but it also enables us to account for macroeconomic dynamics in time-fixed effect models. This was especially important in this study because we focus on the period of financial crisis and recession during which SNAP participation, out-of-pocket household health expenditures, and household financial strain showed simultaneous increases. Fourth, the PSID offers a rich set of control variables. In addition to demographic and income-related variables, the health files of the PSID consist of an exhaustive list of health conditions of the household members.

We excluded the households whose primary respondents were 65 or older, because the relationship between program participation, health expenditures, and financial ratios for older Americans can be quite different than other age groups. Despite the importance of health expenditures in their budget, many older households have access to benefits that are out of reach to younger households, such as Medicare and Social Security. Moreover, the elderly households start receiving distributions from their retirement savings and pension plans, making their financial ratios interpretation different from those of working-age counterparts. After exclusions, the sample consisted of 133,418 household-year observations.

*Variables*

The dependent variable in this study was a dichotomous measure of whether the household participated in the SNAP in the given year. This was measured by the question in the PSID "Did you or anyone else in your family receive food stamp benefits at any time last year?" Those who gave an affirmative answer were considered SNAP participants.

In this study, household out-of-pocket medical expenditures were defined to include insurance premiums as well as other medical bill payments. Insurance premiums were measured as the total health insurance premiums paid by the household for the past two years either directly or through automatic deductions. The past two years' out-of-pocket expenses on nursing home and hospital bills, doctor, outpatient surgery, dental bills, prescriptions, in-home medical care, special facilities, and other services were added. Two variables were created from this sum: first, a logarithm of total out-of-pocket medical expenditures in the previous survey year, and second, a logarithm of the increase in the out-of-pocket medical expenditures since the previous survey year.

The household liquidity constraint was measured using the liquidity ratio. Following Grafova (2011), we considered liquid assets exceeding six months' worth of income as a healthy liquid asset ratio, and therefore defined the liquidity ratio as the total household liquid assets divided by six months' household income. Liquid assets included funds in checking and savings accounts, money market funds, certificate of deposit, government savings bonds, treasury bills, shares of stock in publicly held corporations, mutual funds, or investment trusts that are not employer-based pensions or IRAs. When the respondents were unable to specify the actual amount (fewer than 2% of the respondents), the PSID imputed the values. A higher liquidity ratio was considered to indicate a less constrained household finance.

This study controlled for income and income drop, current and past health conditions, insurance coverage, demographic characteristics, and state and year effects. First, the controls for health conditions included self-reported health status and health deterioration of the household head and spouse, severe conditions such as stroke, heart attack, and lung conditions, chronic conditions such as diabetes, arthritis, blood pressure, and mental health problems of the head and the spouse. Public and private health insurance coverages were also controlled for. The regression model controlled for health variables from the current survey period as well as from the previous survey period. Second, the demographic controls included age, gender, race and ethnicity, education, number of children, marital status, employment status, vehicle ownership, and home ownership. In the prior literature, home and vehicle ownerships have been found to be significant predictors of food access and food consumption (Fitzpatrick and Ver Ploeg 2010; Guo 2011). We also controlled for the region of residence (Northeast, Mid-Central, South, and West as defined by the Census Bureau). Third, income variables included a logarithm of family income, and dichotomous variables indicating whether or not income dropped since the last survey year, and whether the household receives Temporary Assistance for Needy Families (TANF). Fourth, state-level variations in policy environment relating to health care and SNAP rules were controlled through state fixed effects. Year-to-year dynamics in macro-level correlates of SNAP caseload were controlled through year fixed effects.

*Regression Models*

Suppose *Y* is the latent variable for the likelihood of SNAP participation, *MedExp* is the total out-of-pocket medical expenditure, **Liquidity** is a vector of liquidity constraint, *Income* is the total household income, *IncomeDrop* is a dichotomous indicator that the household income dropped since last survey year, **X** is a vector of demographic controls, and **H** is a vector of heath condition controls. The probability of the *i*-th household in state *s* participating in the SNAP at time *t* can be written as:

$$Y_{ist} = \beta_0 + \beta_1 \ln(MedExp_{ist-1}) + \beta_2(\Delta \ln MedExp_{ist}) + \boldsymbol{\beta_3 Liquidity_{ist-1}} + \boldsymbol{\beta_4 Liquidity_{ist}}$$
$$+ \beta_5 \ln(Income_{ist-1}) + \beta_6 IncomeDrop_{ist} + \boldsymbol{\beta_7 X_{ist}} + \boldsymbol{\beta_8 H_{ist-1}} + \boldsymbol{\beta_9 H_{ist}}$$
$$+ \boldsymbol{\gamma_s} + \boldsymbol{\delta_t} + \varepsilon_{ist}$$

Where, $\Delta \ln MedExp_t = \ln(MedExp_t/MedExp_{t-1})$-1, $\boldsymbol{\gamma}$ and $\boldsymbol{\delta}$ are state and year fixed effects, respectively, and $\varepsilon$ is the regression residual. The lagged medical expenditure variable is included to measure the size effect, and the difference of logs of the medical expenses is included in the model to measure the percentage change in medical expenditure. The coefficients $\beta$ were estimated in maximum likelihood fixed effect Logit.

Because previous research indicated that households were inadequately prepared to deal with the sudden out-of-pocket medical expenses arising from health shocks (McIntyre et al. 2006; Feenberg and Skinner 1992), we expect $\beta_2$ to be positive. Also, previous studies found that liquidity constraint was negatively associated with financial well-being of households (Grafova 2011; Grenninger et al. 1996; Cox and Jappelli 1993), therefore the coefficient vectors $\boldsymbol{\beta_{3, 4}}$ are predicted to be negative for the liquidity ratio.

One of the research questions in this study is whether the effect of increased medical expenditures on SNAP participation can be minimized if the household is not liquidity constrained and can therefore borrow or draw from own savings. Liquidity-constrained households would be more likely to seek public assistance programs following high medical expenditures than households with sufficient liquid assets. Previous evidence suggests that out-of-pocket medical expenditures can lead to financial strain especially among households with inadequate reserves of emergency funds (Kim et al. 2012; Kim and Lyons 2008). However, very little research has been done to examine the interaction between medical expenditures and liquidity constraint in affecting program participation. Therefore we estimated a model:

$$Y_{ist} = \beta_0 + \beta_1 \ln(MedExp_{ist-1}) + \beta_2 \Delta(lnMedExp_{ist}) + \beta_3 \boldsymbol{Liquidity_{ist-1}} \ln(MedExp_{ist-1})$$
$$+ \beta_4 \boldsymbol{Liquidity_{ist-1}} \Delta(lnMedExp_{ist}) + \beta_5 \ln(Income_{ist-1})$$
$$+ \beta_6 IncomeDrop_{ist} + \boldsymbol{\beta_7 X_{ist}} + \boldsymbol{\beta_8 H_{ist-1}} + \boldsymbol{\beta_9 H_{ist}} + \boldsymbol{\gamma_s} + \boldsymbol{\delta_t} + \varepsilon_{ist}$$

and expect $\beta_3$ and $\beta_4$ to be negative for the liquidity ratio. Negative coefficients for the interaction between the liquidity ratio and medical expenditure would mean that a household's prior liquid asset holdings make medical expenses less difficult to deal with so they might be able to do without the SNAP. In other words, medical expenses can be more detrimental to the SNAP caseload if households already have little savings.

**Results**

*Descriptive Statistics*

The results from the descriptive statistics are presented in Table 1. The results indicate that SNAP participation among the respondents during 2003-2011 has ranged from 16.5% to 16.8%; SNAP participation was the highest in the 2009 wave. In this dataset, 46% of the respondents were white, and 21% had an educational attainment of college or higher. Approximately 75% of the respondents were employed and the average household income for the population ranged from $67,854 in 2003 to $62,674 in 2011. Household income, which is adjusted in 2003 dollars, showed a declining trend over the five waves of this data. During this period, 45% of the respondents held adequate liquid assets. Approximately 36% of the respondents were renters, and 75% owned a car. Eighty-nine percent of the respondents had either private or public health insurance coverage in 2003; however, the participation rate in health insurance declined to 72% in 2011. Interestingly, the average out-of-pocket medical cost (adjusted in 2003 dollars) was $9,069 in 2003, and increased to a peak of $13,926 in 2009 and $12,832 in 2011. Approximately a quarter of the respondents self-reported being in excellent health during this period, and approximately 18% reported having health problems that limited their ability to work.

The results from the t-tests that compare the characteristics of SNAP participants with then non-SNAP participants are reported in Table 2. The results indicate that when compared with respondents who did not participate in the SNAP program ($1,068), the SNAP participants had a significantly higher increase in out-of-pocket medical costs ($1,395) during the 2003-2011 periods. The results also indicate that a higher percentage of SNAP participants (4.80%) self-reported being in poor health when compared with the non-SNAP participants (4.30%). SNAP participants reported a higher rate (18.3%) of having health conditions that limited their ability to work. A comparison of the financial ratios indicates that a substantially lower percentage of SNAP participants (18%) had an adequate liquidity ratio when compared with the non-participants (48%) during this period. The average income of the SNAP participating households ($22,869) was significantly lower than the average income of the non-SNAP participating ($69,429) households. Interestingly, as many as 63.6% of the SNAP participating households reported a drop in income during the 2003-2011 periods, but only 45% of the non-SNAP participating households did so.

*Likelihood of participation in SNAP*

The results from Table 3 indicate that the increase in medical expenses since the previous period *(Odds=1.01; p<0.001)* and the level of medical expenditure in the previous period *(Odds=1.012; p<0.001)* were both significant and positively associated with SNAP participation when controlling for income, change in income, health-related factors, and state and yearly fixed effects. Interestingly, participants with income drop since the previous period had approximately 3 times (*Odds=3.711; p<0.001*) the likelihood of SNAP participation when compared with households who did not experience an income drop since the previous period. Conversely, income and private health insurance participation were negatively associated with SNAP participation. In addition to these, the liquidity ratio was negatively associated with SNAP participation after we included the financial ratio controls in the second model. The final model from Table 3 adds demographic controls to the previous model that included income, income drop, financial ratios, health factors, and insurance coverage . The results indicated that the increase in medical expenditure and the level of medical expenditure in the previous period were both positively associated with SNAP participation. Similarly, income drop was also positively associated with SNAP participation. Conversely, the liquidity ratio, income, and participation in private health insurance coverage were negatively associated with SNAP participation among the respondents. Among the demographic variables, age and gender (female) were significant and positively associated with SNAP participation. Black households were significantly more likely to participate in SNAP than other racial groups. Similarly, when compared with respondents with an educational attainment of graduate school or higher, the respondents with educational attainment of lower than college were more likely to participate in SNAP. The likelihood of SNAP participation also increased with the number of children in the household and with receiving TANF. Similarly, when compared with those who were married, women who were divorced, widowed, or never married were more likely to participate in SNAP. Conversely, the working respondents were significantly less likely to participate in SNAP, whereas when compared with homeowners, the respondents who were renters were more likely to participate in SNAP.

*Likelihood of SNAP participation for households with income lower than 185% of the poverty line*

The results from Table 4 indicate that after controlling for state and yearly fixed effects, income, income drop, and health controls, the increase in medical expenses (*Odds=1.008; p<0.001*) and medical expenditure (*Odds=1.012; p<0.001*) in the previous period were positively associated with SNAP participation. After adding financial ratios and insurance coverage controls, the results indicate that the increase in medical expenses and medical expenditure in the previous period were still positively associated with SNAP participation, while the likelihood of SNAP participation was negatively associated with the liquidity ratios of households in the current and previous period. The increases in medical expenditure, and medical expenditure in the previous period were positively associated with SNAP participation. Conversely, the liquidity ratios in the current and previous periods were negatively associated with SNAP participation after the demographic variables were added as control variables in the model.

*Examination of the Interaction between increase in health expenditure and the liquidity ratio*

The interaction between the increase in medical expenditure and the liquidity ratio is examined in the model presented in Table 5. The results indicate that after controlling for state and yearly fixed effects, income, income drop, and health related variables, the increases in medical expenses (*Odds=1.006; p<0.001*) and medical expenditure in the previous period (*Odds=1.008; p<0.01*) were positively associated with SNAP participation. Furthermore, the interaction between the liquidity ratio and the increase in medical expenses (*Odds=0.221; p<0.001*) was negatively associated with SNAP participation. The significance of this interaction term indicates that having an adequate liquidity ratio relieves the effect of the increased medical expenditure. The interaction term remained significant when the lagged liquidity ratio and health insurance coverage variables were included in the second model. Increase in medical expenditure was positively associated with SNAP participation in the full model (third) (*Odds=1.005; p<0.01*) after inclusion of demographic controls, but medical expenditure in the previous period was no longer significant. Conversely, the interaction variable for liquidity ratio and medical expenses (*Odds=0.618; p<0.05*), and liquidity ratio *Odds=0.412; p<0.001*) in the previous period were negatively associated with SNAP participation in the full model.

**Discussion**

Findings from the present study reveal that the health care burden of households may contribute to whether or not these households participate in SNAP. Increases in health care costs were positively associated with SNAP participation in the entire sample, as well as for the low-income households (<185% FPL). Additionally, the average out-of-pocket health care spending of SNAP participants was not significantly different from that of non-participants. But, despite that the average household income of non- participants was approximately three times higher than that of SNAP participants. Another interesting takeaway from the findings of this study was that the out-of-pocket medical expenses increased at a much higher rate for SNAP participants than for non-participants. This disparity suggests a higher burden of medical expenditures on low-to-moderate income households (Bernad et al. 2014; Selden 2009). This finding is aligned with previous findings that health care expenditures increased with financial strains (Purtell et al. 2012) and financial instability among households. A substantial increase in medical expenses may lead to an increase in SNAP participation among households lacking in sufficient savings to buffer against excessive strain in household consumption.

Financial assets and savings may be used to smooth consumption and reduce SNAP participation. The liquidity ratio was negatively associated with SNAP participation, consistent with previous research on liquidity constraint and household financial well-being (Grafova 2011; Grenninger et al. 1996; Cox and Jappelli 1993). Liquidity-constrained households were more likely to participate in SNAP upon health shocks. More importantly, findings support the importance of savings for low-income groups. Low-income households need to be encouraged to establish buffer savings. Additionally, there is a need for policies and financial products that can lower the barriers to participation and provide better access to financial services for lower income households (Sherraden 2013). Interestingly, having an adequate liquidity ratio in the previous period was also significant and negatively associated with SNAP participation among low income (<185% FLP) households. The interaction between household financial stability and health care burden was significant. Financial assets can be used to alleviate the health care burden and may reduce SNAP participation. Having inadequate reserves of emergency funds in dealing with health shocks can lead to financial strain in households (Kim et al. 2012; Kim and Lyons 2008). Although, this study examined the dynamics of medical expenditure and SNAP participation, some inaccuracies may be present due to self-reporting of the SNAP participation and medical expenditure variables (Kreider et al. 2012).

**Implications**

This study provides valuable insights into the issue of financial burden of medical care and its relationships to SNAP participation. Further research is needed to examine the effects of the Patient Protection and Affordable Care Act (ACA) of 2010 on health care burden, household finances, and SNAP participation. With the ACA, more households would have access to health insurance and the program would help pay the medical costs that often distress peoples' personal finances.  It may be possible to observe the effects of the decrease in health care costs on SNAP participation. A recent paper that examined the effects of the health care reform in Massachusetts found broader impacts on household finances beyond health and health care utilization such as the total amount of debt, credit scores, and personal bankruptcies (Mazumder and Miller 2014). The impacts of the ACA Medicaid expansion on financial burden from medical spending among the low-income Americans will be uneven from state to state due to the differences in states' participation in the program (Caswell, Waidmann, and Blumberg 2013). Many of states that currently have not adopted the Medicaid expansion have traditionally high SNAP participation (Kaiser Family Foundation 2015).

This study also calls for additional research on the role of household financial stability in SNAP participation. Health shock is one of the causes for financial instability of households. However, researchers have argued the importance of savings and asset building to prevent financial crisis and suggest that even households with very limited resources can still build savings (Schreiner and Sherraden 2005).

The current study offers important implications for policies and programs. Our findings suggest that reducing the health care burden of households may not only improve health outcomes but also decrease SNAP participation. Better coordination of public assistance such as food and health programs may be associated with reductions in SNAP caseload. Our study suggests it may be crucial to investigate whether coordination of the two programs may increase efficiency in public finance and government budgets. Further, health care burden with or without insurance can be financially draining, especially for households with chronic health conditions. Currently, in most states medical care costs are deductible expenses for households with members who are elderly and for households with disabilities in calculating SNAP eligibility and benefits. Eligible households can deduct out-of-pocket medical expenses that are more than $35. The deduction for excessive medical expenses can lead to a substantial increase in SNAP benefits (USDA, 2014)**.** A broad array of medical expenses, including transportation costs to a pharmacy or a doctor's office, over-the-counter drugs, medical supplies, and home renovations to increase accessibility, are eligible for deduction**.** For eligible seniors or disabled individuals, a claim of $50-$200 in monthly medical expenses can result in a monthly increase of $7-$69 in SNAP benefits. Although claiming the medical expenses deduction for SNAP eligibility can be

beneficial for the eligible individuals, extant research shows that this deduction is underutilized (Jones, 2014). Jones (2014) found that about 12 percent of the eligible households actually claim this deduction, and suspected that the actual number of eligible households could be much larger. Policies that can promote and educate eligible households of the range of available deductible expenses, so that they can better leverage the benefits, should be made a priority for inclusion in the SNAP education programs that are administered across the country. An additional policy suggestion is to extend the excess medical deduction when calculating SNAP eligibility to households younger than 60 as well. Further research may be necessary to determine the costs and benefits of such a policy. Given the findings of this study, it is possible the potential benefits to household well-being may well outweigh the costs of extending such excess medical deductions to all SNAP eligible households, especially for households who need continuous health care spending for chronic health conditions.

The findings from this study also suggest a need to revisit the current policy on asset limits for SNAP eligibility. Presently, the federal asset limit for SNAP benefits is set at $2,000 (or $3,250 if the household has an elderly or disabled person). While many states have increased the limit or eliminated it through a broad-based categorical eligibility, the current federal limit does not encourage sufficient precautionary savings for financial emergencies such as unexpected medical expenses. In fact, inadequate asset limits under the current policy might create a disincentive to build savings and financial security for low-income families, the lack of which could lead to SNAP participation.

**Table 1. Descriptive Statistics**

| | Variable | 2003 N=26,675 | | 2005 N=26,686 | | 2007 N=26,689 | | 2009 N=26,675 | | 2011 N=26,693 | |
|---|---|---|---|---|---|---|---|---|---|---|---|
| | | %, Mean | Freq. | % | Freq. | % | Freq. | % | Freq. | % | Freq. |
| Dependent | Food stamps recipient | 16.53% | 4409 | 16.56% | 4419 | 16.77% | 4476 | 16.80% | 4481 | 16.5% | 4417 |
| | | | | | | | | | | | |
| Demographic | White | 46% | 12271 | 46% | 12276 | 46% | 12277 | 46% | 12271 | 46% | 12279 |
| | Black | 37% | 9870 | 37% | 9874 | 37% | 9875 | 37% | 9870 | 37% | 9876 |
| | Hispanic | 9% | 2401 | 9% | 2402 | 9% | 2402 | 9% | 2401 | 9% | 2402 |
| | Other | 8% | 2134 | 8% | 2135 | 8% | 2135 | 8% | 2134 | 8% | 2135 |
| | Female | 27% | 7202 | 27% | 7205 | 27% | 7206 | 27% | 7202 | 27% | 7207 |
| | Age | 42 | | 42 | | 43 | | 43 | | 44 | |
| | Education | | | | | | | | | | |
| | <HS | 21% | 5602 | 21% | 5604 | 21% | 5605 | 21% | 5602 | 21% | 5606 |
| | High school | 33% | 8803 | 33% | 8806 | 33% | 8807 | 33% | 8803 | 33% | 8809 |
| | Some college | 25% | 6669 | 25% | 6672 | 25% | 6672 | 25% | 6669 | 25% | 6673 |
| | College | 14% | 3735 | 14% | 3736 | 14% | 3736 | 14% | 3735 | 14% | 3737 |
| | Graduate school | 7% | 1867 | 7% | 1868 | 7% | 1868 | 7% | 1867 | 7% | 1869 |
| | Number of kids | 1.3 | | 1.3 | | 1.3 | 34696 | 1.3 | | 1.3 | |
| | Married | 57% | 15205 | 57% | 14944 | 57% | 15213 | 57% | 14938 | 58% | 15482 |
| | Never married | 22% | 5869 | 22% | 5871 | 22% | 5872 | 19% | 5068 | 21% | 5606 |
| | Widowed | 4% | 1067 | 4% | 1067 | 4% | 1068 | 3% | 800 | 4% | 1068 |
| | Divorced | 17% | 4535 | 18% | 4803 | 17% | 4537 | 21% | 5869 | 17% | 4538 |

Health Care Expenditures and SNAP

| | | N=26675 | | N=26686 | | N=26689 | | N=26675 | | N=26693 | |
|---|---|---|---|---|---|---|---|---|---|---|---|
| | | %, Mean | Freq. | % | Freq. | % | Freq. | % | Freq. | % | Freq. |
| Region | North central | 25% | 6669 | 25% | 6672 | 25% | 6672 | 25% | 6669 | 25% | 6673 |
| | South | 43% | 11470 | 43% | 11475 | 43% | 11476 | 43% | 11470 | 43% | 11478 |
| | West | 18% | 4802 | 18% | 4803 | 18% | 4804 | 18% | 4802 | 19% | 5072 |
| | Other region | 1% | 267 | 1% | 267 | 1% | 267 | 1% | 267 | 0% | 0 |
| Socioeconomic | Employed | 75% | 20006 | 75% | 20015 | 76% | 20284 | 75% | 20006 | 75% | 20020 |
| | Income | $67,854 | | $67,559 | | $67,639 | | $64,074 | | $62,674 | |
| | Receive TANF | 2.22% | 592 | 2.37% | 632 | 2.32% | 620 | 2.22% | 589 | 2.27% | 606 |
| Financial | Have Emergency Funds | 26% | 6936 | 26% | 6938 | 26% | 6939 | 26% | 6936 | 26% | 6940 |
| | Have Liquidity | 45% | 12004 | 45% | 12009 | 44% | 11743 | 44% | 11737 | 44% | 11745 |
| | Own a car | 72% | 19206 | 72% | 19214 | 72% | 19216 | 72% | 19206 | 73% | 19486 |
| | Renter | 36% | 9603 | 36% | 9607 | 36% | 9608 | 36% | 9603 | 36% | 9609 |
| Health | Have Health Insurance | 89% | 23741 | 86% | 22950 | 82% | 21889 | 78% | 20807 | 72% | 19219 |
| | Total OOP cost | $9,069 | | $11,049 | | $12,389 | | $13,926 | | $12,832 | |
| | Increase in OOP costs | | | $2,281 | | $1,596 | | $1,818 | | ($553) | |
| Family Health | Family Poor Health | 4% | 1067 | 5% | 1334 | 4% | 1068 | 4% | 1067 | 4% | 1068 |
| | Family Health Conditions | 24% | 6402 | 26% | 6938 | 26% | 6939 | 27% | 7202 | 27% | 7207 |
| | Health Limits Work | 19% | 5068 | 18% | 4803 | 18% | 4804 | 17% | 4535 | 17% | 4538 |
| | Family psych Problem | 5% | 1114 | 5% | 1334 | 6% | 1601 | 7% | 1867 | 7% | 1869 |

**Table 2. T-Test comparisons for SNAP Participants**

|  |  | SNAP=1 | SNAP=0 | T test P-value |
|---|---|---|---|---|
| Family Health | Total Out-of-Pocket Cost | $11,903 | $12,185 |  |
|  | Increase in Out-of-Pocket Cost | $1,395 | $1,068 | * |
|  | Family Health Condition | 26% | 26% |  |
|  | Family Psychological Issues | 6% | 6% |  |
|  | Poor Perceived Health | 4.80% | 4.30% | *** |
|  | Health Limits Work | 18.30% | 17.80% | * |
| Financial | Liquidity Ratio (Fin. Assets/6 Months Income)>=1 | 18% | 48% | *** |
|  | Income | $22,869 | $69,429 | *** |
|  | Fall in Income | 63.60% | 45.00% | *** |

**Table 3. Logit model for the likelihood of SNAP participation (n=133,418)**

| | SNA Participation | (1) Coeff. | Odds | (2) Coeff. | Odds | (3) Coeff. | Odds |
|---|---|---|---|---|---|---|---|
| | Change in medical expenditure | 0.008*** | 1.01 | 0.008** | 1.015 | 0.005** | 1.006 |
| | Med expenditure (t-1) | 0.011*** | 1.012 | 0.011*** | 1.177 | 0.013** | 1.014 |
| | Liquidity ratio | | | -1.461*** | 0.529 | -0.839*** | 0.41 |
| | Liquidity ratio (t-1) | | | -0.663 | 0.324 | -0.178 | 0.731 |
| | Log Income | -0.978*** | 0.183 | -0.933*** | 0.178 | -0.882*** | 0.398 |
| | Income Drop | 0.795*** | 3.711 | 0.721*** | 3.561 | 0.671*** | 2.062 |
| Family Health | Perceived Poor Health | 0.28 | 1.146 | 0.264 | 1.264 | 0.251 | 1.149 |
| | Chronic Health Conditions | 0.261 | 1.128 | 0.215 | 1.233 | 0.209 | 1.199 |
| | Health Limits Work | 0.241 | 0.109 | 0.194 | 1.207 | 0.14 | 1.134 |
| | Family Psychological Issues | -0.126 | 0.782 | -0.087 | 0.894 | -0.074 | 0.918 |
| | Have Private Health Ins | -0.234*** | 0.768 | -0.217*** | 0.764 | -0.154** | 0.856 |
| | Have Public Health Ins. | -0.031 | 0.968 | -0.028 | 0.873 | -0.045 | 0.958 |
| Demographic | White | | | | | 0.045 | 1.046 |
| | Black | | | | | 0.838*** | 2.311 |
| | Hispanic | | | | | 0.192 | 1.212 |
| | Female | | | | | 0.634*** | 1.883 |
| | Age | | | | | 0.494*** | 1.419 |
| | Education | | | | | | |
| | LtHS | | | | | 2.134*** | 4.234 |
| | High school | | | | | 1.773*** | 3.844 |
| | Some college | | | | | 1.404*** | 3.198 |
| | College | | | | | 0.669 | 1.735 |
| | Number of kids | | | | | 0.544*** | 1.529 |
| | Never married | | | | | 0.435*** | 1.459 |
| | Widowed | | | | | 0.417*** | 1.433 |

| | | | | | | |
|---|---|---|---|---|---|---|
| Socioeconomic | Divorced | | | | 0.582*** | 1.597 |
| | Employed | | | | -1.196*** | 0.324 |
| | Receive TANF | | | | 0.339*** | 1.382 |
| | Own a car | | | | -0.249 | 0.794 |
| | Renter | | | | 0.802*** | 2.321 |
| | State / Yearly FE | YES | | YES | YES | |

(Logit coefficients and odds ratio, ***p<.001, **p<.01, *p<.05)

**Table 4. Logit model for the likelihood of SNAP participation when income < 185% of Poverty (n=39,659)**

| SNAP Participation | (1) Coeff. | Odds | (2) Coeff. | Odds | (3) Coeff. | Odds |
|---|---|---|---|---|---|---|
| Increase in medical expenditure | 0.008* | 1.032 | 0.007* | 1.007 | 0.007* | 1.008 |
| Med expenditure (t-1) | 0.012*** | 1.003 | 0.011** | 1.012 | 0.011** | 1.011 |
| Liquidity ratio | | | -1.331*** | 0.264 | -0.843*** | 0.431 |
| Liquidity ratio (t-1) | | | -0.373*** | 0.732 | -0.339*** | 0.713 |
| State & year fixed effects | Yes | | Yes | | Yes | |
| Income and income drop | Yes | | Yes | | Yes | |
| Health variable controls | Yes | | Yes | | Yes | |
| Insurance coverage controls | No | | Yes | | Yes | |
| Demographic controls | No | | No | | Yes | |

(Logit coefficients and odds ratio, ***p<.001, **p<.01, *p<.05)

**Table 5. Logit model for the likelihood of SNAP participation with Increase in medical expenditure and liquidity ratio interaction (n=133,418)**

| SNAP Participation | (1) Coeff. | Odds | (2) Coeff. | Odds | (3) Coeff. | Odds |
|---|---|---|---|---|---|---|
| Increase in medical expenditure | 0.008*** | 1.006 | 0.008*** | 1.006 | 0.007** | 1.005 |
| Med expenditure (t-1) | 0.011** | 1.008 | 0.009** | 1.007 | 0.012 | 1.012 |
| Liquidity*Increase in medical expenses | -1.460*** | 0.221 | -1.457*** | 0.233 | -0.885*** | 0.412 |
| Liquidity ratio (t-1) | | | -0.484** | 0.696 | -0.461* | 0.618 |
| State & year fixed effects | Yes | | Yes | | Yes | |
| Income and income drop | Yes | | Yes | | Yes | |
| Health variable controls | Yes | | Yes | | Yes | |
| Insurance coverage controls | No | | Yes | | Yes | |
| Demographic controls | No | | No | | Yes | |

(Logit coefficients and odds ratio, ***p<.001, **p<.01, *p<.05)